\documentclass[onecolumn,superscriptaddress]{revtex4}

\usepackage{braket}
\usepackage{graphicx,float}
\usepackage{epsfig}
\usepackage{dcolumn}
\usepackage{ulem}
\usepackage{bm}
\usepackage{amsmath}
\usepackage{amssymb}
\usepackage{hyperref}
\hypersetup{colorlinks=true,linkcolor=blue,citecolor=blue,urlcolor=black}

\newcommand{\ii}{\text{i}}
\newcommand{\sy}{\text{s}}
\newcommand{\tot}{\text{tot}}
\newcommand{\pt}{\text{PT}}
\newcommand{\apt}{\text{APT}}

\begin{document}
	\title{Recurrence Theorem for Open Quantum Systems}

\author{Zhihang Liu}
\affiliation{Department of Physics, College of Science, North China University
	of Technology, Beijing 100144,  China.}

\author{Chao Zheng}
\email{czheng@ncut.edu.cn}  
\affiliation{Department of Physics, College of Science, North China University
	of Technology, Beijing 100144, China.}
\affiliation{School of Energy Storage Science and Engineering, North China University of Technology, Beijing 100144, China.}
\affiliation{Beijing Laboratory of New Energy Storage Technology, Beijing 100144, China}

\date{\today}

	\begin{abstract}
	Quantum (Poincaré) recurrence theorem are known for closed quantum (classical) systems. Can recurrence happen in open systems?  We provide the  recurrence theorem for  open quantum systems  via non-Hermitian (NH) description. We find  that PT symmetry and  pseudo-Hermitian symmetry protect recurrence for  NH  open quantum  systems and the recurrence fails with the symmetry breaking. 
	Applying  our theorem to   PT-symmetric systems,  we reveal why  quantum recurrence happens in PT-unbroken phase but fails in PT-broken phase, which was misunderstood before.
	A contradiction emerges when we apply our theorem to anti-PT symmetric systems and we settle it,  revealing that distinguishability   and von Neumann entropy are generally  not effective to describe the  information dynamics in NH  systems.
	A new approach  is developed to investigate   the  information dynamics of  NH  systems. For anti-PT symmetric systems in PT-broken phase, we find  there are three information-dynamics patterns:  oscillations with an overall decrease (increase) , and periodic oscillations.  The periodic oscillations (information complete retrieval) happen only if the spectrum of NH Hamiltonian is real. The three patterns  degenerate to the periodic oscillation  using distinguishability  or von Neumann entropy because normalization of non-unitary evolved states leads to   loss of information. We conclude with a discussion of the physical meaning behind the recurrence in open systems and give the direction of recurrence theorem not limited to conservative systems in classical mechanics. 
\end{abstract}

\maketitle

\section{Introduction}

For a closed quantum system governed by a time-independent Hamiltonian with discrete energy eigenvalues, quantum recurrence theorem \cite{bocchieri1957quantum} states that  the system will return  to a state arbitrarily close to its initial state after a finite time. Recent researches about recurrence in quantum systems are limited to closed quantum system \cite{wallace2015recurrence, grunbaum2013recurrence}.
However, no quantum system is completely isolated from its environment and  open quantum systems is ubiquitous in nature. Is there a quantum recurrence theorem for open quantum systems? We provide one in this paper via non-Hermitian (NH) description of open quantum systems.

NH operators  as an effective description of open quantum systems dates back to 
the studies of  open quantum systems by  Feshbach. The radiative decay in reactive nucleus has been analyzed by an effective NH Hamiltonian associated with the decay of the norm of a quantum state, indicating the presence of nonzero probability flow to the outside of nucleus \cite{feshbach1958unified, feshbach1962unified, ashida2020non}. 

In recent decades, NH physics attracts growing interesting both in theory and experiments.  On one hand, NH  physics with parity-time symmetry can be seen as a complex extension of the conventional quantum mechanics \cite{bender2002complex}, having novel properties. On the other hand, it closely related to open and dissipative systems of realistic physics \cite{breuer2002theory, barreiro2011open, hu2020quantum, del2020driven, zheng2021universal, zhang2024localization}.  Typical NH systems feature different symmetries, such as the parity-time (PT) symmetry \cite{bender1998real, kawabata2017information, zheng2013observation, PhysRevLett.132.083801},  anti-PT  symmetry \cite{yang2017anti, li2019anti, peng2016anti, bergman2021observation, yang2020unconventional, choi2018observation, zheng2019duality, wen2020observation}, pseudo-Hermitian symmetry \cite{mostafazadeh2002pseudo, mostafazadeh2004pseudounitary, jin2022unitary, xu2023pseudo}, anyonic-PT symmetry  \cite{zheng2022quantum, longhi2019anyonic, arwas2022anyonic}. 
While it is physically understandable that there will be no recurrence in open systems interacting with environment in a random way,
we find  that PT symmetry and  pseudo-Hermitian symmetry protect recurrence for  NH open quantum systems. 
Study of quantum information  \cite{bennett2000quantum, xavier2020quantum},  
such as entropy \cite{bianchi2018linear, bian2020quantum, ding2022information}, entanglement  \cite{vidmar2017entanglement, fang2022entanglement}, and critical phenomena  \cite{kawabata2017information, xiao2019observation, vidmar2018volume} is fundamental  for    quantum computing and communication.  We show the application prospects of our theorem in the theory of quantum information by applying it to typical NH systems. Since quantum (Poincaré) recurrence theorem are only applicable to closed quantum (classical) systems,  the quantum recurrence theorem for open quantum systems  thus  provides a deeper understanding of recurrence in physical systems, such as that judicious balancing of gain and loss   is underlying the recurrence in   PT-symmetric systems.  Inspired by our theorem, 
we expect  a   recurrence theorem not limited to  conservative systems in classical mechanics.  

%

This paper is structured as follows. 
In section \uppercase\expandafter{\romannumeral2}, we proof the quantum recurrence theorem for open quantum systems (in this paper, we refer to it as our theorem).
Applying  our theorem to   PT-symmetric systems,  we reveal why  quantum recurrence happens in PT-unbroken phase but fails at PT-broken phase, which was misunderstood before.
A contradiction emerges when we apply our theorem to anti-PT symmetric systems and we settle it,
revealing that distinguishability  (trace distance) adopted in \cite{kawabata2017information, wen2020observation, xiao2019observation} and von Neumann entropy adopted in  \cite{ding2022information, bian2020quantum} are generally  not effective to describe the  information dynamics in NH open quantum systems.
A new approach  based on non-normalized density matrix $\Omega(t)$ is developed to investigate   the  information dynamics of  NH  systems and we find it is highly correlated with NH entropy \cite{sergi2016quantum}. For anti-PT symmetric systems in PT-broken phase,  there can be  three information-dynamics patterns:  oscillation with an overall decrease (increase) , and periodic oscillation.  The periodic oscillation (information complete retrieval) happens only if the spectrum of NH Hamiltonian is real. The three patterns will degenerate to the periodic oscillation  using distinguishability  (trace distance) or von Neumann entropy because normalization of non-unitary evolved states leads to  loss of information about the NH systems. We conclude with a discussion of the physical meaning behind the recurrence in open systems and give the direction of recurrence theorem not limited to conservative systems in classical mechanics.

\section{ recurrence theorem for open quantum systems}

For an open quantum system governed by Hamiltonian $H_{\sy}(t)$, 
which is NH, arbitrary dimensional and time-dependent in general, we can always extend  $H_{\sy}(t)$ to a Hermitian one by introducing an ancilla qubit \cite{wu2019observation}.  
With the state evolution of NH system satisfying (in units $\hbar=1$)
\begin{equation}
	\ii \frac{d}{dt}|\psi(t)\rangle=H_{\sy}(t)|\psi(t)\rangle ,
\end{equation}
the state evolution of the extended Hermitian system satisfies
\begin{equation}
	\label{Eq2}
	i\frac{d}{dt}|\Psi(t)\rangle=H_{\tot}(t)|\Psi(t)\rangle,
\end{equation}
with
\begin{equation}
	\label{Eq3}
	\begin{aligned}
		|\Psi(t)\rangle&=|\psi(t)\rangle|-\rangle+|\chi(t)\rangle|+\rangle .
	\end{aligned}
\end{equation}
When a measurement is applied on the ancilla qubit and $|-\rangle$ is postselected, the system evolution  governed by the NH Hamiltonian $H_{\sy}(t)$ can be produced \cite{ozawa1984quantum}.
If $H_{\tot}$ is time-independent, we can apply the  quantum recurrence theorem for closed quantum systems to it and we know the extended Hermitian system  returns to a state arbitrarily close to its initial state after a finite time, and the NH system obtained by the projection of the extended Hermitian system also returns to a state arbitrarily close to its initial state after a finite time.
However, in general, $H_{\tot}$ is time-dependent, even though $H_{\sy}$ is time-independent. 
The method to get $H_{\tot}(t)$  \cite{wu2019observation}  and its proof is pretty complex, but for our purpose, we will focus on $M(t)$, because we find that the time-dependence of $H_{\tot}$ can come solely from $M(t)$, i.e., if $M(t)$ can be time-independent, so does $H_{\tot}(t)$.
For time-independent arbitrary dimensional $H_{\sy}$ with  discrete spectrum, 
\begin{equation}
	M(t)=e^{-\ii H_{\sy}^{\dagger}t } M(0) e^{\ii H_{\sy}t },
\end{equation}
\begin{equation}
	H_{\sy}^{\dagger} \ket{\phi_{n}}=\kappa_{n}\ket{\phi_{n}}. \label{s1}
\end{equation}
$M(0)$   is chosen to ensure that $M(t)-I$ keeps positive for all $t$.
It turns out that we can choose $M(0)$ as the linear combination of $ \ket{\phi_{n}}  \bra{\phi_{n}} $  with appropriate coefficients (with $M^{\prime}(0)$ being a positive operator, of which all the eigenvalues are larger than 1, $M(0)$=$M^{\prime}(0) / \mu$ for some $\mu>0$ is what we need), then 
\begin{equation}
	M(t)=\sum_{n=1}^{\infty} \alpha_{n} M_{n}(t) \quad (\alpha_{n}>1),
\end{equation}
\begin{equation}
	\begin{aligned}
		M_{n}(t)
		&=e^{-\ii H_{\sy}^{\dagger}t } \ket{\phi_{n}}  \bra{\phi_{n}} e^{\ii H_{\sy}t }\\
		&=e^{-\ii (\kappa_{n}-\kappa_{n}^{\ast})t}  \ket{\phi_{n}}  \bra{\phi_{n}} ,\\
	\end{aligned}
\end{equation}
where $\ast$ means complex conjugate.
So, if the spectrum of  $H_{\sy}^{\dagger}$ is real, $\kappa_{n}-\kappa_{n}^{\ast}=0$ and  $M(t) $ is time-independent. And thus $H_{\tot}(t)$ can be time-independent, which indicates that quantum recurrence will happen in the NH system governed by $H_{\sy}$.
With
\begin{equation}
	H_{\sy} \ket{\nu_{n}}=\nu_{n}\ket{\nu_{n}}, \label{s2}
\end{equation}
\begin{equation}
	\bra{\nu_{n}}H_{\sy}^{\dagger} =\nu_{n}^{\ast}\bra{\nu_{n}}, \label{s3}
\end{equation}
since the spectrum  of   $H_{\sy}$ is  unique, comparing Eq (\ref{s1}) and Eq (\ref{s3}),  we have \cite{faisal1981time, wong1967results}
\begin{equation}
	\nu_{n}^{\ast}=\kappa_{n}
\end{equation}
So,  the spectrum  of   $H_{\sy}^{\dagger}$ is  real if and only if the spectrum  of   $H_{\sy}$ is  real.
We arrive at the quantum recurrence theorem for open quantum systems. $\\$

$ \mathbf{Quantum \; Recurrence \; Theorem:}$ For an open quantum system governed by a time-independent  Hamiltonian  $H_{\sy}$ with discrete spectrum, if the spectrum of  $H_{\sy}$ is real, the  system  returns to a state arbitrarily close to its initial state after a finite time.   $\\$

Clearly, the quantum recurrence theorem for closed  quantum systems can be seen as the restriction of our theorem to closed systems.
For our theorem to be useful, 
we look for NH Hamiltonian with real spectrum and that is why PT symmetry and pseudo-Hermitian are important. 
If  $H_{\sy}$  has a discrete spectrum and admits a complete set of biorthonormal eigenvectors, the spectrum of $H_{\sy}$  is real if and only if there is an invertible linear operator $O$ such that $H_{\sy}$  is $OO^{\dagger}$-pseudo-Hermitian \cite{mostafazadeh2002pseudo2}.  
Generic  finite dimensional NH Hamiltonian $H_{\sy}$   admit a  complete set of biorthonormal eigenvectors if it is away from  exceptional degeneracy (not at exceptional point).  The existence of complete set of biorthonormal eigenvectors  for infinite-dimensional Hilbert space is not known \cite{brody2013biorthogonal, bergholtz2021exceptional}. Thus, the best we can say is the spectrum of  finite dimensional NH $H_{\sy}$ is real if and only if it is $OO^{\dagger}$-pseudo-Hermitian and not at exceptional point. However,  infinite dimensional PT-symmetric  Hamiltonian can have real spectrum and  
many    PT-symmetric  Hamiltonians are rigorously proved to have entirely real and positive spectra \cite{dorey2001spectral, dorey2001supersymmetry}. 
For a   PT-symmetric system, the Hamiltonian $H_{\pt}$ satisfies $[{\rm PT}, H_{\pt}]=0$. It is in PT-unbroken phase if each eigenstate of Hamiltonian  is simultaneously the eigenstate of the  PT operator, in which case the entire spectrum is real.  Otherwise, it is in PT symmetry  broken phase, and some pairs of eigenvalues become complex conjugate to each other. By our theorem, we now understand why  quantum recurrence happens in PT unbroken phase but fails in PT broken phase, which was misunderstood before for finite dimensional PT-symmetric system and led to wrong statements \cite{kawabata2017information}. The PT symmetry protects the recurrence, which fails with the symmetry breaking. It is  worth to remark that 
the recurrence is not necessarily  periodic,  i.e.,  the second recurrence period does not need to be same as  the first recurrence period. 
The argument about  the periodic oscillation of distinguishability  in \cite{kawabata2017information} is only reasonable for finite-dimensional systems and  in a not rigorous way. 
$\\$

\section{Quantum information oscillation}
By adopting distinguishability or von Neumann entropy to characterize the information dynamics of the systems \cite{wen2020observation, ding2022information}, it is argued that quantum information complete retrieval can happen in the broken phase of  two-level anti-PT symmetric systems with complex eigenvalues. 
But by  our theorem, we know 
there will be no quantum recurrence, in broken phase or unbroken phase, and  quantum information complete retrieval should not happen. Furthermore, for arbitrary dimensional  anti-PT symmetric systems, in broken phase or unbroken phase, there will be no quantum recurrence for the reason below.
Mathematically, a conventional   PT-symmetric Hamiltonian would become anti-PT symmetric on multiplying by imaginary unit $\ii$, implying properties of anti-PT symmetric systems conjugate to those of PT systems \cite{peng2016anti}. So, generally, the eigenvalues of a anti-PT symmetric Hamiltonian are pure imaginary numbers in  unbroken phase and complex numbers in broken phase. Applying our theorem, we know there will be no quantum recurrence in anti-PT symmetric systems unless 
some special conditions are satisfied and thus the spectrum of the  anti-PT symmetric Hamiltonian is real. 

Here comes the contradiction. In \cite{wen2020observation}, quantum information complete retrieval   happens whenever the two-level anti-PT symmetric system is   in PT symmetry broken phase.
The problems is that  distinguishability (trace distance) and  von Neumann entropy are generally  not effective  describing the quantum information dynamics of NH open quantum systems,  because  both deal solely with normalized quantum state, i.e., require the conservation of probability, which is against the nature of NH systems and leads to loss of information. Recall that the distinguishability of two quantum states is \cite{nielsen2010quantum, kawabata2017information}. 
\begin{gather}
	D \left( \rho_{1} \left( t \right), \rho_{2} \left( t \right) \right)
	= \frac{1}{2} \mathrm{Tr} \left|\,\rho_{1} \left( t \right) - \rho_{2} \left( t \right)\,\right| ,
\end{gather}
where $| \rho | := \sqrt{\rho^{\dag} \rho}$,  $\rho_{1, 2}$ are normalized density matrices and $\mathrm{Tr} $ indicates trace of density matrix. In Hermitian quantum mechanics, with initial wave function normalized,  the Schrödinger equation  automatically preserves the normalization of the wave function.  However, for  the investigation of   NH open quantum systems  in  \cite{kawabata2017information, wen2020observation, ding2022information, fang2022entanglement, fang2021experimental},   with  normalized initial quantum states, their  normalization   is artificial rather then  intrinsic to the Schrödinger equation.  
We argue that the non-normalized density matrix  is  essential to characterize the information dynamics in NH systems.  It is ubiquitous in nature that the probability in an open quantum system  effectively becomes non-conserved due to the flows of energy, particles, and information between the system and the external environment \cite{ashida2020non}.   
In the study of  radiative decay in reactive nucleus, which is  analyzed by an effective NH Hamiltonian, the essential idea is that  the decay of the norm of a quantum state  indicates the presence of nonzero probability flow to the outside of nucleus \cite{feshbach1958unified, feshbach1962unified, ashida2020non}. The non-conserved norm indicates there is information flow between the NH system and environment. Thus, the non-conserved norm is essential for describing information dynamics in NH systems.

We develop a new approach to characterize the information dynamics in NH systems. 
Density matrix $\Omega (t)$ undergoing no artificial normalization is essential.
Boltzmann's  entropy formula and Shannon's  entropy formula state  the logarithmic connection between entropy and probability. We borrow this wisdom and take the natural logarithm of ${\rm Tr }\,\Omega (t)$. 
We  find that $-\ln {\rm Tr }\,\Omega (t)$ can serve as a new description for the information dynamics in NH systems. 
Furthermore,  non-Hermitian entropy \cite{sergi2016quantum, li2022non}
is found to be  highly correlated with $-\ln {\rm Tr}(\Omega(t))$.  
In units Boltzmann constant $k_{B} =1$, non-Hermitian entropy is defined as:
\begin{equation}
	S_{{\rm NH}}(t)=-{\rm Tr}\,[\rho (t)\ln\Omega (t)].
	\label{NH}
\end{equation}
Comparing to von Neumann entropy $S_{{\rm vN}}$:
\begin{equation}
	S_{{\rm vN}}(t)=-{\rm Tr}\, [\rho (t)\ln \rho (t)].
	\label{vn}
\end{equation}
the only difference between the expressions of $S_{{\rm NH}}(t)$ and $S_{{\rm vN}}(t)$ is the use of $\Omega (t)$.
We employ the usual Hilbert-Schmidt inner product  when  we investigate the effective non-unitary dynamics of an open quantum system described by NH Hamiltonians \cite{kawabata2017information, brody2016consistency, brody2012mixed}.
 The dynamics governed by $H_{\sy }$  
is described by   
\begin{equation}
	\Omega (t)=e^{-\ii H_{\sy } t } \, \Omega (0) \, e^{\ii H_{\sy } ^{\dagger }t}  \label{dy}
\end{equation}
\begin{equation}
	\rho(t)=\Omega(t)/{\rm Tr}(\Omega(t)) \label{normal},
\end{equation}
For arbitrary dimensional ${H}_{\sy}$ with     eigenenergies $E_{n} + \ii \Gamma_{n}$, 
\begin{equation}
	H_{\sy} \ket{\varphi_{n}} = \left( E_{n} + \ii \Gamma_{n} \right) \ket{\varphi_{n}},
\end{equation}
with $  \braket{\varphi_{n} | \varphi_{n} }=1$. 
Conduct the spectral decomposition of the NH dynamics given by Eq.(\ref{dy}):
\begin{equation}
	\Omega \left( t \right)
	= \sum_{ij} \Omega_{ij} e^{-\ii \left( E_{i} - E_{j} \right) t} e^{\left( \Gamma_{i} + \Gamma_{j} \right) t} \ket{\varphi_{i}} \bra{\varphi_{j}}  
	\qquad  \Omega_{ij} = \frac{\braket{\chi_{i} |\,\Omega(0)\,| \chi_{j}}}{\braket{\chi_{i} | \varphi_{i}} \braket{\varphi_{j} | \chi_{j}}} \; ,
\end{equation} 
with  $\chi_{i}$ being the left eigenstates of $H_{\sy}$  \cite{brody2013biorthogonal, kawabata2017information}.
Non-conserved norm of the quantum state origins from the the  decay parts $e^{ \Gamma_{n}  t} $ (complex spectrum) and the non-orthogonality of the eigenstates $\ket{\varphi_{n}} $ of NH ${H}_{\sy}$.
The decay parts $e^{ \Gamma_{n}  t} $ disappear if we normalize the state,  which leads to wrong  characterization of the NH system.  
Define the eigenstates with the largest (second largest) imaginary part as $ \ket{\varphi_{1}}  $ ( $\ket{\varphi_{2}}  $). After a sufficiently long time, $ \ket{\varphi_{1}}  $ and $\ket{\varphi_{2}}  $ dominate the dynamics.  
With arbitrary initial state $\ket{\varphi_{0}}= { \sum_{n=1}}  c_{n} \ket{\varphi_{n}}  $  ($\Omega(0)= \ket{\varphi_{0}} \bra{\varphi_{0}}$), the NH dynamics asymptotically behaves as
\begin{equation}
	\begin{aligned}
		&	\Omega \left( t \right)
		\sim  |c_{1}| ^{2}  \, e^{2\Gamma_{1}t} \ket{\varphi_{1}} \bra{\varphi_{1}}+ |c_{2}| ^{2} \,  e^{2\Gamma_{2}t} \ket{\varphi_{2}} \bra{\varphi_{2}}+ (  c_{1} c_{2}^{\dagger} \,  e^{-\ii (E_{1} - E_{2})t} \ket{\varphi_{1}} \bra{\varphi_{2}}  + {\rm H.c.}  )  e^{(\Gamma_{1}+\Gamma_{2} ) t}  \label{ome} ,
	\end{aligned}
	\end{equation}
	and we have 
	\begin{equation}
	\begin{aligned}
		&-\ln  {\rm Tr} \, \Omega(t) \sim   -\ln [\,  |c_{1}| ^{2}  \, e^{2\Gamma_{1}t}+ |c_{2}| ^{2} \,  e^{2\Gamma_{2}t}  +  (  c_{1} c_{2}^{\dagger} \,  e^{-\ii (E_{1} - E_{2})t} \braket{\varphi_{2} |\varphi_{1}}  + {\rm c.c.}  )  e^{(\Gamma_{1}+\Gamma_{2} ) t}  \, ]   \label{-ln}
		\; .
	\end{aligned}
	\end{equation}
	By Eq.(\ref{-ln}),   three  information-dynamics patterns are possible:  oscillation with an overall decrease (increase) , and periodic oscillation.  The periodic oscillation (information complete retrieval)   happens only if the spectrum of  ${H}_{\sy}$ is real and thus there is no decay parts, which usually happens  for  
	PT-symmetric systems in  unbroken phase.
Distinguishability and von Neumann entropy   can describe the phenomenon of  information complete retrieval  for   PT-symmetric systems in  unbroken phase \cite{kawabata2017information, bian2020quantum, xiao2019observation}  because the   spectra are real and thus there is no decay parts before or after normalization of quantum states. For anti-PT symmetric systems (and other NH systems), spectra are complex  in general and normalization causes the loss of information about the NH systems, which leads to contradictory results to our theorem.

\begin{figure}[t]
	\includegraphics[width=\linewidth]{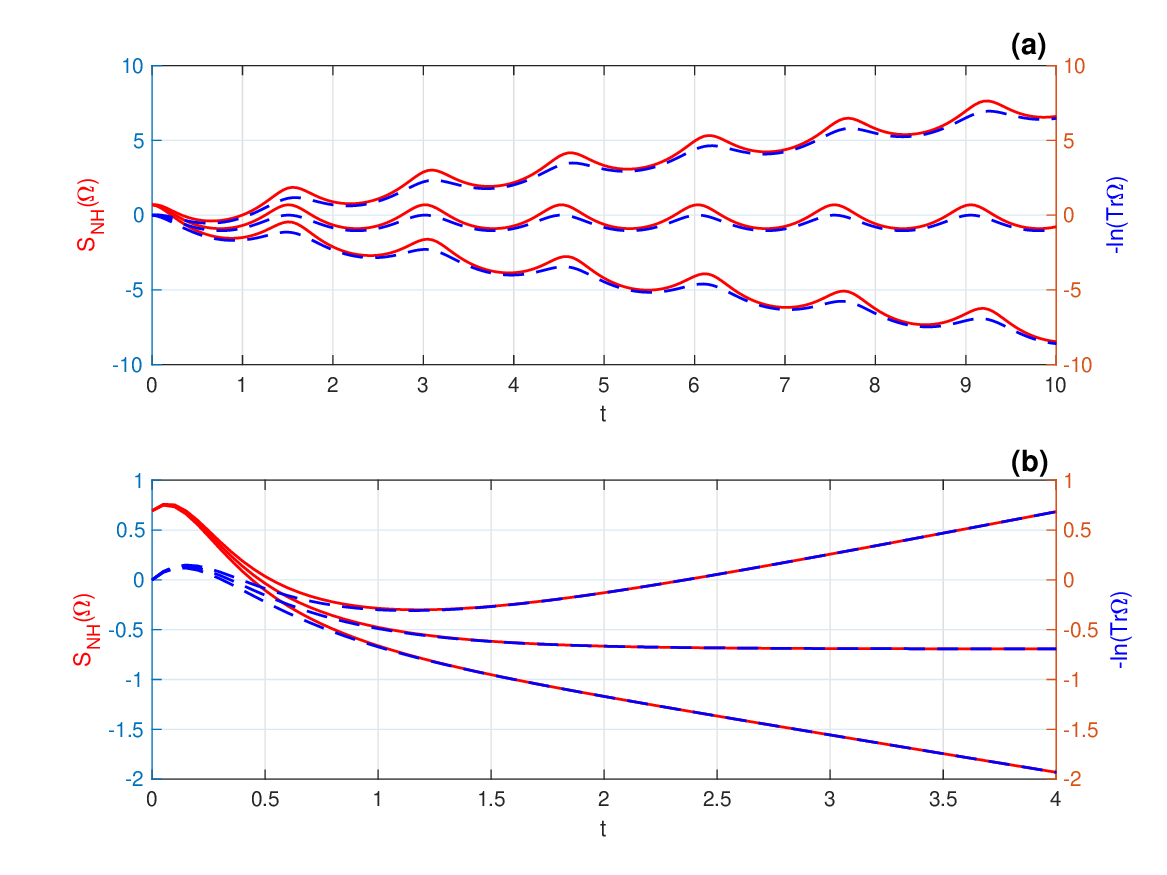}  
	\caption{For all  figures in this paper, the red line represents the non-Hermitian entropy $S_{{\rm NH}} \,\Omega (t)$ (the left vertical axis) ,  the  dashed blue line represents the $-\ln {{\rm Tr}(\Omega(t) )} $ (the right vertical axis),  and the scales of  left and right vertical axes are set to be same. 
		Parameter setting for $H_{\apt}$:  $r_{1}=2$,  $\theta_{1} =\pi/6$. 
		(\textbf{a}) $r=\frac{5}{\sqrt{3}}$,  $\delta<0$, $H_{\apt}$ in PT  symmetry broken phase.
		$\theta=11\pi/ 24, \;  \pi/2, \;  13\pi/24$, corresponding to $r\cos \theta>0,\; =0, \; <0$,  respectively.
		The three information-dynamics patterns:  oscillation with an overall decrease (increase) , and periodic oscillation are well predicted by Eq.(\ref{Adel-0}).  
		(\textbf{b})  $\theta=2\pi/3$, $\delta>0$, $H_{\apt}$ in PT  symmetry unbroken phase.
		\label{APT3}  }
\end{figure}   

\begin{figure}[t]
	\includegraphics[width=\linewidth]{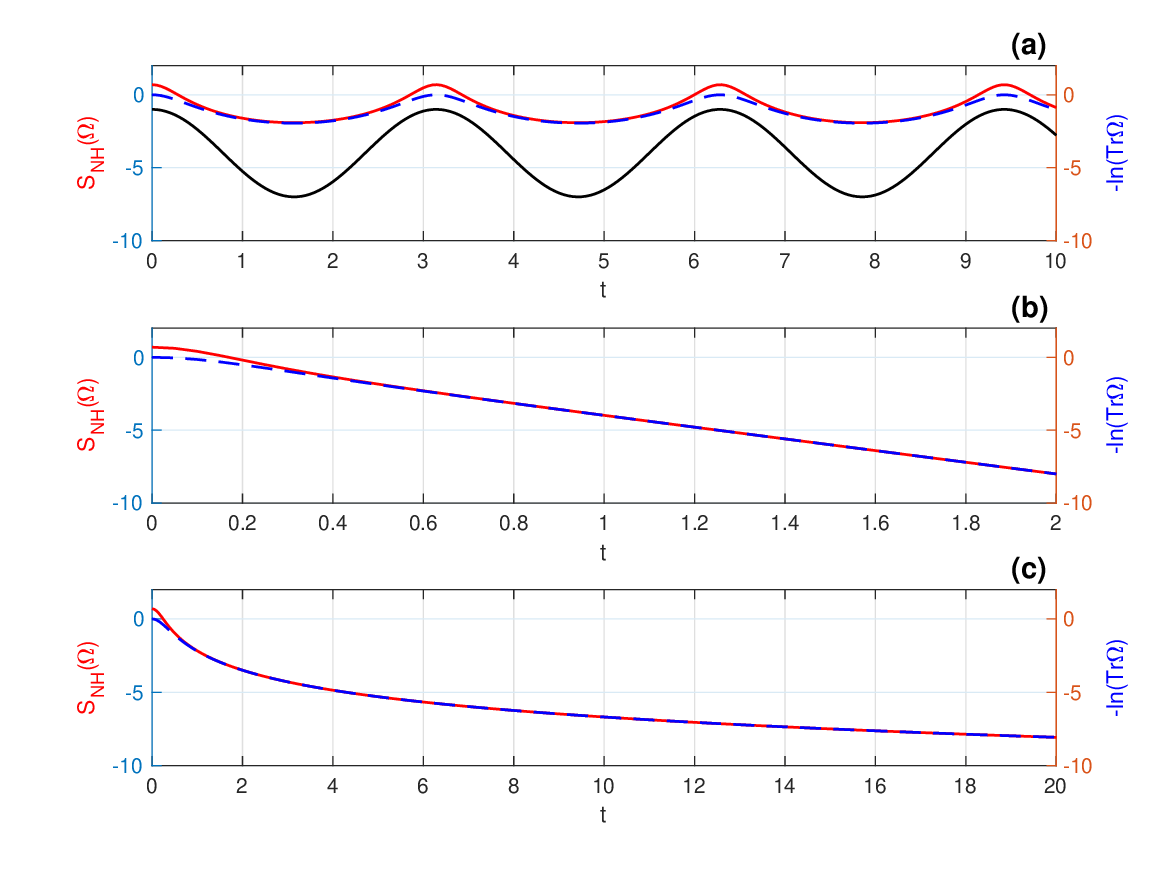} 
	\caption{ parameter setting for $H_{\pt}$:  $r_{1}=2$, $\theta_{1} =\pi/6$.
		(\textbf{a})  $r=2\sqrt{3}$, $\delta =1 $, $H_{\pt} $ in  PT symmetry  unbroken phase. The black line represents  $-{\rm Tr}\, \Omega(t) $, which shows why we adopt $-\ln {\rm Tr}\, \Omega(t)$.
		(\textbf{b})  $r=4\sqrt{2}$, $\delta =-4$,  $H_{\pt} $ in   PT symmetry broken phase.
		(\textbf{c})  $r=4$, $\delta =0 $,   $H_{\pt} $ at exceptional point.
		\label{pt3}   } 
\end{figure}

\begin{figure}[t]
	\includegraphics[width=\linewidth]{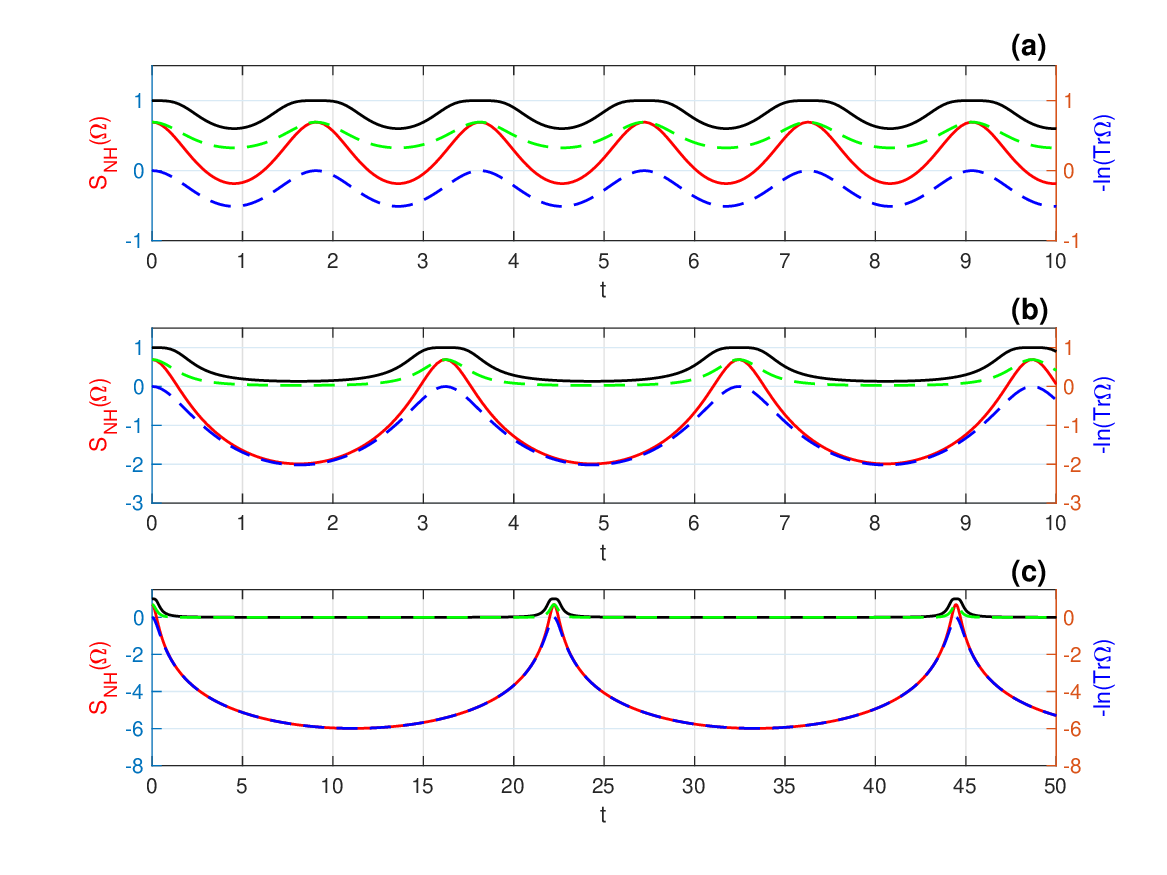}  
	\caption{ The black line represents distinguishability $D$, the dashed green line represents von Neumann entropy $S_{{\rm vN}}  $.  
		Parameter setting:  $r_{1}=2$, $\theta_{1} =\pi/6$, $\theta =\pi/6$.  
		(\textbf{a})  $r=2$,  $\delta =3 $;
		(\textbf{b})  $r=3.5$, $\delta =\frac{15}{16} $;
		(\textbf{c})  $r=3.99$, $\delta =0.02  $, system in the vicinity of exceptional point. 
		\label{4d}     }
\end{figure}

\subsection*{Two-level Systems}
As  a proof-of-principle example, we consider generic two-level anti-PT symmetric system governed by  $H_{\apt}$. With the parity operator $\mathcal{P}$  given by 
$\begin{pmatrix} 0&1\\1&0\end{pmatrix}$, and $\mathcal{T}$ being the operation of complex conjugation, $H_{\apt}$   can be expressed as a four-parameter family of matrices: 
\begin{equation}
	H_{\apt} = \ii
	\begin{pmatrix}
		re^{\ii\theta }   & r_{1} e^{\ii \theta_{1}  }\\ r_{1} e^{-\ii\theta_{1}} & re^{-\ii\theta }
	\end{pmatrix} \; ,	
\end{equation}
where $r, \theta, r_{1}, \theta_{1}$  are   real.
A family of corresponding   PT-symmetric $H_{\pt}$ is 
\begin{equation}
	H_{\pt} =
	\begin{pmatrix}
		re^{\ii \theta }   & r_{1} e^{\ii\theta_{1}  }\\ r_{1} e^{-\ii\theta_{1}} & re^{-\ii\theta } 
	\end{pmatrix}. \label{PT}
\end{equation}
The energy eigenvalues of $H_{\apt}$ are
\begin{equation}
	E_{\pm } =\ii r\cos \theta \pm \ii \sqrt{\delta }  \; ,
\end{equation}
with
\begin{equation}
	\delta =r_{1} ^{2} -r^{2} \sin ^{2} \theta \; .
\end{equation}
For   $H_{\apt}$,  when  $\delta >0$,  the system in  PT symmetry unbroken  phase;  when  $\delta <0$,  the system  in  PT symmetry broken phase;  the exceptional point of $H_{\apt}$    locates  at $\delta =0$.
The dynamics governed by  $H_{\apt}$ is described by   
\begin{equation}
	\Omega (t)=e^{-\ii H_{\apt } \, t}  \, \Omega (0) \, e^{\ii H_{\apt }^{\dagger } \, t} \; , 
\end{equation}
For  $H_{\apt}$ in PT symmetry unbroken phase, with initial  state being  a maximally mixed state  in two-level system, i.e.,  $	\Omega (0)=\frac{1}{2} I  $ ($I$ being the identity operator), 
\begin{equation}
	\begin{aligned}
		&-\ln \, 	{\rm Tr}\,\Omega(t)
		= -2tr\cos \theta 
		-\ln (\frac{1+a}{2}\cosh  2\sqrt{\delta }t /\hbar+\frac{1-a}{2})  \; ,
	\end{aligned} 	\label{unAPT}
\end{equation}
where $a=\frac{r_{1}^2+r^2\sin ^{2}\theta }{\delta}  \ge 1$. 
For  $H_{\apt}$ in PT symmetry broken phase, 
\begin{equation}
	\begin{aligned}
		&	-\ln \, 	{\rm Tr}\,\Omega(t)
		=  -2tr\cos \theta  
		-\ln (   \frac{1-b}{2} \cos (2\sqrt{-\delta } t/\hbar)+\frac{1+b}{2}    ) \; ,
	\end{aligned}  
	\label{Adel-0}
\end{equation}
where $b=\frac{r_{1}^2+r^2\sin ^{2}\theta }{-\delta} \ge 1 $.

As we show in FIG.$\ref{APT3}$(a), For $H_{\apt}$ in PT symmetry broken phase, there are  three information-dynamics patterns:  oscillation with an overall decrease (increase) , and periodic oscillation.  The periodic oscillation  happens only if the spectrum of  ${H}_{\apt}$ is real, in this case, $r \cos \theta =0$.
For $H_{\apt}$  in PT symmetry unbroken phase, there can be three information-dynamics patterns too, i.e.,  decrease, increase and asymptotically stable, as we show in FIG.$\ref{APT3}$(b). 
The three patterns in broken (unbroken) phase will degenerate to the periodic oscillation (asymptotically stable) using von Neumann entropy or distinguishability \cite{wen2020observation, ding2022information}, The degenerate results are wrong in general.  Furthermore, FIG.$\ref{APT3}$ shows that  $S_{{\rm NH}}(t)$
and $	-\ln \, 	{\rm Tr}\,\Omega(t)$ are highly correlated.  The correlation between the two description is also demonstrated in other figures of this paper.

The reader may still have reservation about the effectiveness of $S_{{\rm NH}}  $ and $-\ln {\rm Tr } \, \Omega(t)$. 
We further demonstrate it  by adopting them to characterize the  information dynamics of the generic two-level   PT-symmetric systems governed by $H_{\pt}$ in Eq.(\ref{PT}). 
For $H_{\pt}$  in PT unbroken phase, 
\begin{equation}
	\begin{aligned}
		-\ln \, 	{\rm Tr}\,\Omega(t)=  
		-\ln ( \frac{1-a}{2} \cos 2\sqrt{\delta }t/\hbar+\frac{1+a}{2} )
	\end{aligned} 	\label{unPT}, 
\end{equation}	
where $a=\frac{r_{1}^2+r^2\sin ^{2}\theta }{\delta}  \ge 1$. 
By Eq.(\ref{unPT}), the information dynamics is periodic oscillation (information complete retrieval) in the   unbroken phase of  $H_{\pt}$. 
The result is showed in FIG.(\ref{pt3})(a). 
Furthermore, we compare the four kinds of information dynamics descriptions: distinguishability (trace distance) $D$, von Neumann entropy $S_{{\rm vN}}  $, $S_{{\rm NH}}  $ and $-\ln {\rm Tr }(\Omega)$.  
$D$ and $S_{{\rm vN}}  $ deal only with normalized density matrix $\rho(t)$, while $S_{{\rm NH}}  $ and $-\ln {\rm Tr }(\Omega)$ make use of non-normalized $\Omega(t)$.   $S_{{\rm vN}}  $ and $S_{{\rm NH}}  $ characterize  information dynamics with the concept of entropy, while $D$ and $-\ln {\rm Tr }(\Omega)$ do it with trace.
To make the four kinds of descriptions comparable, the initial states are chose so that they all start from their maximum value:  distinguishability $D$ is maximal when the two quantum states are orthogonal, so the initial states $\rho_{1}(0)$ and $\rho_{2}(0)$  are set to be  $|0\rangle\langle0| $ and $|1\rangle\langle1|$, respectively;  the initial states of  the other three are all set to be the maximally mixed state $\Omega(0)= \frac{1}{2} I$.  The  four kinds of information dynamics descriptions of  $H_{\pt} $  are shown in FIG.(\ref{4d}).
We remark that non-unitary evolution governed by a NH Hamiltonian provides a non-zero von Neumann  entropy production. 
Distinguishability $D$ and $S_{{\rm vN}}  $  are highly correlated 
and the correlation  becomes stronger as the system getting closer to exceptional point, just as $S_{{\rm NH}}  $ and $-\ln {\rm Tr }(\Omega)$ do.  
If we use  $\frac{\partial}{ \partial t} D $ or  $\frac{\partial}{ \partial t} S_{{\rm vN}} $ to indicate the information flow and thus to characterize the non-Markovian and Markovian processes of the    system,   FIG.(\ref{4d}) shows that  when the systems get closer to exceptional point, $\frac{\partial}{ \partial t} D $ and  $\frac{\partial}{ \partial t} S_{{\rm vN}} $  are practically zero for most part of every period,  indicating a dynamics  neither non-Markovian nor Markovian.
In contrast,   $S_{{\rm NH}}  $ and $-\ln {\rm Tr }(\Omega)$ properly  characterize the non-Markovian and Markovian processes of the  PT-symmetric system.

\section{Conclusion and outlook}
Time evolution governed by NH Hamiltonian is non-unitary and thus the norm of a quantum state is not conserved, indicating the presence of nonzero probability flow between the system and environment. Starting from our theorem, we demonstrate that the widely adopted methods — distinguishability and von Neumann entropy,  are generally  not effective to describe the  information dynamics in NH  systems except for those with real spectra.
Borrowing the wisdom of Boltzmann's entropy formula, we propose a new information-dynamics description based on non-normalized density matrix  and  find it is highly correlated with NH entropy $S_{{\rm NH}}  $. 
For anti-PT symmetric systems in PT-broken phase, the three information dynamics patterns we find degenerate to the periodic oscillation using distinguishability or von Neumann entropy. 
The investigation of entanglement dynamics, concurrence and coherence in NH systems \cite{fang2021experimental, fang2022entanglement} faces the same problem about normalization of non-unitary evolved  quantum states, which can lead to wrong results too. 
The concept of PT symmetry spread out to  many  physical fields such as optics and Bose-Einstein condensates, where a judicious balancing of gain and loss constitutes a  PT-symmetric system \cite{konotop2016nonlinear, jing2014pt, miroshnichenko2011nonlinearly}.  We remark that it is the balanced gain and loss makes  PT-symmetric systems have the features  of   conserved systems, such as real spectra and recurrence. Thus, we propose that recurrence can happen in  open classical systems featuring balanced gain and loss.  
Likewise, the role of pseudo Hermitian in classical mechanics deserves attention. 
A pseudo Hermitian quantization scheme is introduced to  
determine the underlying classical system for pseudo-Hermitian quantum systems    \cite{mostafazadeh2004physical}; 
Classical-quantum correspondence for two-level pseudo-Hermitian systems is proposed and analyzed \cite{raimundo2021classical}. Inspired by  the role of PT symmetry and   pseudo Hermitian symmetry played in our theorem,   we expect  a   recurrence theorem not limited to  conservative systems in classical mechanics.  

$\\$

\paragraph*{Acknowledgements}
This work was supported by the National Natural Science Foundation of China (Grant Nos. 12175002, 11705004, 12381240288), the Natural Science Foundation of Beijing (Grant No. 1222020), the Project of Cultivation for Young top-notch Talents of Beijing Municipal Institutions (BPHR202203034). 




\end{document}